\documentclass[%
 aip,
 amsmath,amssymb,
 reprint,%
]{revtex4-1}

\usepackage{graphicx}
\usepackage{dcolumn}
\usepackage{bm}

\usepackage[utf8]{inputenc}
\usepackage[T1]{fontenc}
\usepackage{mathptmx}
\usepackage{etoolbox}

\makeatletter
\def\@email#1#2{%
 \endgroup
 \patchcmd{\titleblock@produce}
  {\frontmatter@RRAPformat}
  {\frontmatter@RRAPformat{\produce@RRAP{*#1\href{mailto:#2}{#2}}}\frontmatter@RRAPformat}
  {}{}
}%
\makeatother
\begin{document}

\preprint{AIP/123-QED}

\title[Machine Learning algorithms for optimization of magnetocaloric effect in all-$d$-metal Heusler alloys]{Machine Learning algorithms for optimization of magnetocaloric effect in all-$d$-metal Heusler alloys}
\author{D.R. Baigutlin}
 \affiliation{Faculty of Physics, Chelyabinsk State University, 454001, Chelyabinsk, Russia}
 \email{d0nik1996@mail.ru}

\author{V.V. Sokolovskiy}%
 \affiliation{Faculty of Physics, Chelyabinsk State University, 454001, Chelyabinsk, Russia}

\author{V.D. Buchelnikov}
  \affiliation{Faculty of Physics, Chelyabinsk State University, 454001, Chelyabinsk, Russia}

\author{S.V. Taskaev}
  \affiliation{Faculty of Physics, Chelyabinsk State University, 454001, Chelyabinsk, Russia}

\date{\today}

\begin{abstract}
This study examines the application of machine learning algorithms, specifically the Random Forest regression model, to optimize the magnetocaloric effect in all-$d$-metal Heusler alloys. The model was trained using descriptors related to the mean properties of individual atoms, the properties of simple compounds in their ground state, and measures of chemical disorder. It demonstrated high accuracy in predicting structural properties, while exhibiting moderate accuracy in predicting magnetic properties.
To identify optimal alloy compositions, a genetic algorithm was used to find those with the greatest differences in magnetization during martensitic transitions. Using this combined approach, the Ni-Co-Mn-Ti alloy system was thoroughly explored, resulting in the discovery of an alloy with a maximum magnetization difference. These results are consistent with previous research based on Density Functional Theory (DFT) and highlight the effectiveness of integrating machine learning with genetic algorithms for the discovery of new materials with outstanding magnetocaloric properties.
The study emphasizes the need for further refinement of models capable of accurately predicting complex magnetic interactions, which is essential for fully leveraging the potential of all-$d$-metal Heusler alloys in practical applications.
\end{abstract}

\maketitle

\section{Introduction}

For an extended period, magnetocaloric cooling has been regarded as an environmentally benign and efficacious alternative to conventional compressor-based cooling technologies~\cite{gschneidner2000magnetocaloric, shen2009recent, sokolovskiy2022review}. This technology has several advantages, including high energy efficiency, the absence of environmentally harmful refrigerants, and relatively low levels of noise and vibration~\cite{kitanovski2020energy}. In recent years, a variety of potential magnetocaloric materials have been identified, including Gd, manganites, Laves phases, Fe-Rh, Mn-As, La-Fe-Si, and Heusler alloys~\cite{gschneidner2000magnetocaloric, shen2009recent}. Additionally, several prototype devices employing magnetocaloric cooling technology and using Gd as the working material have been demonstrated~\cite{kitanovski2020energy, lionte202115kw, lin2024full}. However, for industrial applications of this technology, several challenges must be addressed, such as the cost of the magnetocaloric material and the magnetic system, as well as the frequency of switching between the heat transfer fluid and the working body, as outlined in~\cite{gutfleisch2016mastering}. Heusler alloys are among the most promising candidates because their Curie temperatures can be easily tuned by varying stoichiometric coefficients~\cite{sokolovskiy2015achieving, zhao2017enhanced}, many consist of relatively inexpensive components, and they exhibit some of the highest magnetocaloric effects observed~\cite{liu2012giant, sokolovskiy2015achieving}.

The main issue is thermal hysteresis, which leads to irreversible changes in the magnetocaloric effect during cyclic operation~\cite{gutfleisch2016mastering}. Another challenge, particularly in systems with a giant magnetocaloric effect characterized by a first-order phase transition, is the development of defects under cyclic loading~\cite{gamzatov2018inverse}. These defects cause "smearing" of the transition, reducing the maximum value of the magnetocaloric effect~\cite{waske2015asymmetric}. Therefore, alloys with exceptional mechanical properties are required to solve the problems of rapid heat transfer and resistance to defect formation. However, traditional X$_d$-Y$_d$-Z$_{sp}$ type alloys are highly brittle due to $p-d$ covalent hybridization~\cite{yan2019giant}. A recently proposed solution to this problem is to replace the Z$_{sp}$ element with a transition metal, thereby substituting strong $p-d$ covalent hybridization with metallic bonding between elements. This results in enhanced mechanical properties associated with the martensitic transition~\cite{wei2015realization}. The ductility of all-$d$-metal Heusler alloys, such as Ni$_2$MnTi, significantly surpasses that of classical Ni$_2$Mn(Ga, Al, In, Sn) alloys based on Pettifor's and Pugh's ratios. The magnetocaloric effect values in these alloys are also high and comparable to those of classical Heusler alloys. For instance, Ni$_{35}$Co$_{15}$Mn$_{37}$Ti$_{13}$ exhibits an isothermal entropy change of 20 J kg$^{-1}$ K$^{-1}$ and an adiabatic temperature change of -4 K for a magnetic field change of~2~T~at room temperature.

The number of possible compositions of all-$d$-metal Heusler alloys is vast, and considering that exceptional properties are often found in non-stoichiometric regions, the search space becomes infinite. This makes a classical search method based on a full factorial design, even when incorporating certain physical considerations, very expensive. Identifying suitable materials among numerous possible compositions is a labor-intensive and time-consuming process. 

In recent years, only a limited number of compositions in this series have been studied, most of which are Ni-Mn-Ti~\cite{wei2015realization, yan2019giant}, with the addition of a fourth element such as Co~\cite{wei2015realization, beckmann2023dissipation}, Fe~\cite{zeng2019electronic}, and a fifth element such as Cu~\cite{zhang2022second} and B~\cite{zhang2022reduced}. 
An alternative approach is to use various machine learning algorithms and high-throughput screening. For example, using regression with a Random Forest algorithm, the mechanical properties of all-$d$-metal Heusler alloys, particularly the Pugh ratio, were optimized~\cite{liu2022machine}, and an empirical formula for estimating ductility was proposed.
In the study~\cite{jin2022classifying}, a model was developed to classify cubic and tetragonal phases of the alloy using simple descriptors of $d$-electron occupancy and spin moment with a support vector machine algorithm, achieving a prediction accuracy of 90.4\% relative to DFT. 
In addition to machine learning, high-throughput screening methods are also employed. In the study~\cite{fortunato2024high}, 1881 all-$d$-metal compounds were considered, and by applying constraints on the convex hull energy, magnetic moment magnitude, type of magnetic ordering, and mechanical stability, 11 compounds were identified. The authors of the study~\cite{fortunato2024high} also provided several empirical rules for determining the lattice type based on the electronegativity of the elements and tuning the behavior of the magnetostructural phase transition.

Despite significant progress in using machine learning methods to optimize the properties of Heusler alloys, existing models face several major challenges. One of the key difficulties is accurately predicting the properties of non-stoichiometric compositions, which is critically important for developing new materials with improved properties. The magnetocaloric effect itself is a complex phenomenon, and there are currently no models that can accurately predict all its aspects. While some models successfully predict elastic properties and chemical stability, the challenge of predicting differences in magnetization between phases, Debye temperatures, and other key parameters remains unresolved.

The aim of this study is to develop models for predicting the volume, tetragonality parameter, and magnetization of austenite and martensite phases, as well as an optimization algorithm for the difference in magnetization for a given atomic composition of all-$d$-metal Heusler alloys.

\section{Material and methods}
A random forest regression method was selected for the cell volume, tetragonality parameter and magnetic moment of austenite and martensite. As descriptors $X$=\{$x_1$, $x_2$, ... $x_n$\}, the mean properties of individual atoms and the properties of simple compounds in their ground state were chosen. Shannon entropy was used to estimate chemical disorder by assessing the randomness in atom distribution. For predicting the tetragonality parameter, the alloy volume predicted earlier was used. Similarly, the predictions of magnetic moments incorporated both the previously determined volume and the tetragonality parameter. The complete list of features is provided in Table~\ref{tab:descriptors}. The Python matminer module~\cite{ward2018matminer} was used to calculate all features except for Shannon entropy. 

\begin{table}[htbp]
    \caption{Descriptor identification codes and their full names with physical symbols. Descriptors $x_1-x_8$ refer to mean properties of constituent atoms; $x_9-x_{13}$ refer to mean properties of simple compounds in their ground state; $x_{14}$ describes chemical disorder; and $x_{15}$ and $x_{16}$ pertain to properties of alloys.}
    \begin{ruledtabular}
    \begin{tabular}{cll}
        \textbf{ID} & \textbf{Full Descriptor Name} & \textbf{Symbol} \\
        \hline
        $x_1$  & Mendeleev number & $Z$ \\
        $x_2$  & Atomic weight & $A$ \\
        $x_3$  & Column & $Col$ \\
        $x_4$  & Row & $Row$ \\
        $x_5$  & Electronegativity & $\chi$ \\
        $x_6$  & Number of $d$ valence electrons  & $N_{val}$ \\
        $x_7$  & Number of $d$ unfilled electrons & $N_{unf}$ \\
        $x_8$  & Covalent radius & $r_{cov}$ \\
        $x_9$  & Space group number & $SGN$ \\
        $x_{10}$ & Volume simple compounds & $V^{sc}$ \\
        $x_{11}$ & Bandgap & $E_g^{sc}$ \\
        $x_{12}$ & Magnetic moment & $\mu^{sc}$ \\
        $x_{13}$ & Melting temperature & $T_m^{sc}$ \\
        $x_{14}$ & Shannon entropy & $S$ \\
        $x_{15}$ & Volume & $V$ \\
        $x_{16}$ & Tetragonality ratio & $c/a$ \\
    \end{tabular}
    \end{ruledtabular}
    \label{tab:descriptors}
\end{table}

\begin{figure}
\includegraphics[width=0.49\textwidth]{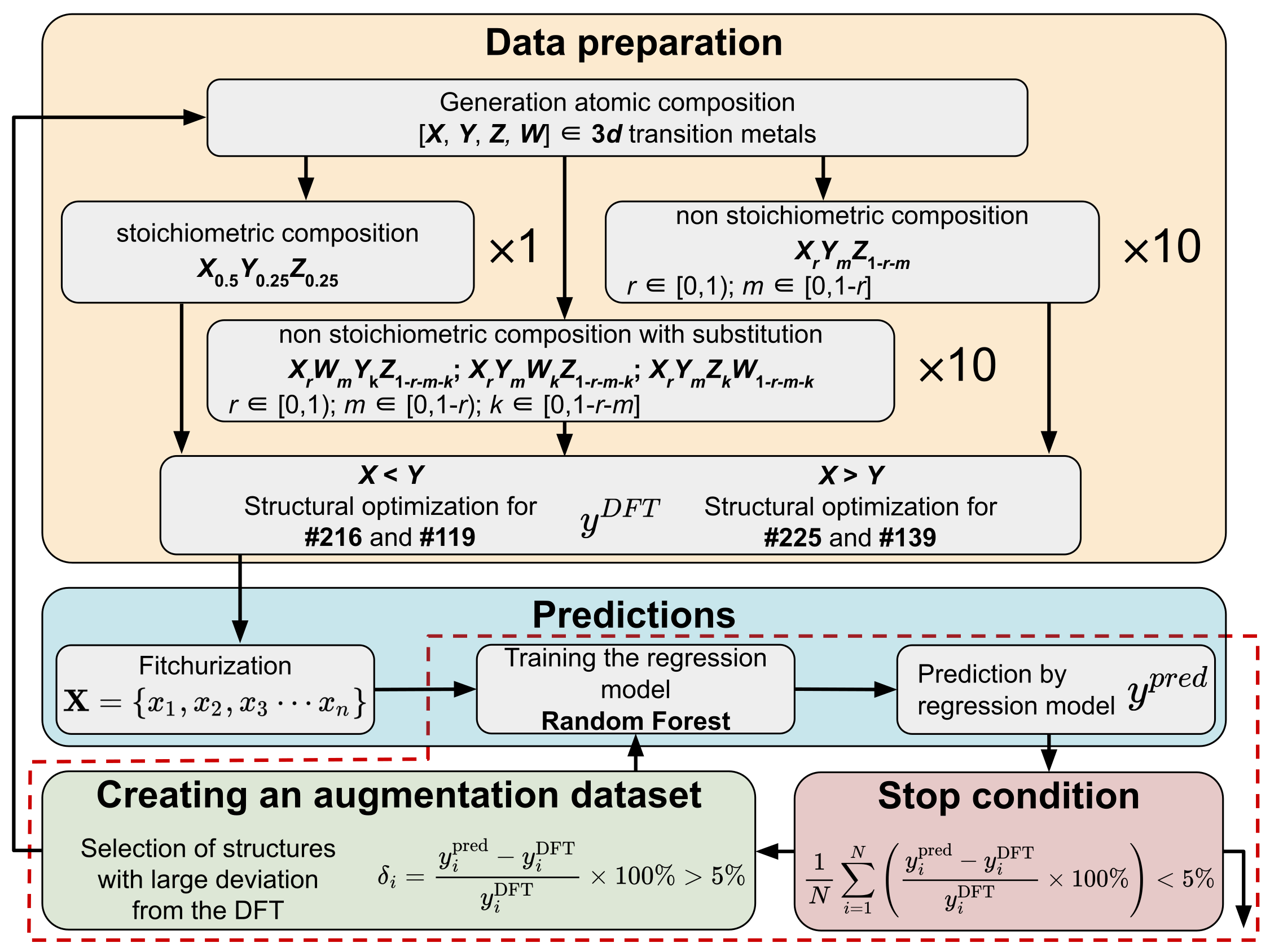}
\caption{The block diagram illustrates the process of constructing a training sample and training a regression model based on active learning. The red dashed line indicates the block of active learning based on relative error.}
\label{block_scheme}
\end{figure}

The training dataset was constructed using an active learning approach, as illustrated in Fig.~\ref{block_scheme}. At this stage, a random atomic composition was generated from four 3$d$-elements [$X, Y, Z, W$]. From this composition, one stoichiometric alloy and ten non-stoichiometric alloys with random coefficients $X_rY_mZ_{1-r-m}$ were created. Additionally, ten compositions were generated with a random concentration of a fourth element at random crystallographic positions: $X_r W_m Y_k Z_{1-r-m-k}$, $X_r Y_m W_k Z_{1-r-m-k}$, and $X_r Y_m Z_k W_{1-r-m-k}$, where $r \in [0, 1]$, $m \in [0, 1-r]$, and $k \in [0, 1-r-m]$. Density Functional Theory (DFT) calculations were performed using the PBE functional~\cite{perdew1996generalized}, implemented in the VASP package~\cite{kresse1996efficient,kresse1999ultrasoft}, for cubic and tetragonal Heusler structures with symmetry groups \#225 (\#139) and \#216 (\#119), involving complete structural optimization. This process provided benchmark values for volume, tetragonality ratio, and magnetic moments of the austenitic and martensitic phases. The choice of structure type, direct \#225 (\#139) or inverse \#216 (\#119), was made according to classical rules for conventional Heusler alloys. It should be noted, however, that this simplification is limited, and for alloys involving all metals, the $Z$ atom must also be considered (see Ref.~\cite{li2023theoretical}).

Subsequently, the generated compositions were described, and a regression model was developed using the Scikit-learn module in Python~\cite{pedregosa2011scikit}. The predictions of the regression model were compared with DFT calculations, and only those compositions for which the deviation between the DFT predictions and the model exceeded 5\% were added to the training dataset. This iterative process continued until, over ten iterations, the average deviation was reduced to less than 5\%.

Cross-validation was used to assess the stability and generalizability of the model. The test set comprised 1,000 samples that were not part of the training set, allowing for an evaluation of the model's performance on a large number of independent data points and its accuracy in predicting the structural and magnetic properties of all-$d$-metal metal Heusler alloys.

Two metrics were employed to determine the accuracy of the regression models in this study. The first metric is the coefficient of determination, $R^2$, which measures the proportion of variance in the dependent variable that can be predicted from the independent variables. The closer this coefficient is to one, the better the model's description.

\begin{equation}
R^2 = 1 - \frac{SS_{res}}{SS_{tot}} = 1 - \frac{\sum_{i=1}^{n} (y_i - \hat{y}_i)^2}{\sum_{i=1}^{n} (y_i - \bar{y})^2},
\end{equation}
where $SS_{res}$ represents the sum of squared residuals and $SS_{tot}$ represents the total sum of squares. Here, $y_i$ and $\hat{y}_i$ denote the actual and predicted values, correspondingly.

The second metric is the root mean square error (RMSE), which quantifies the root of the average squared differences between actual and predicted values. It provides a measure of the model's precision, with lower RMSE values indicating a closer fit of the model to the data points.

\begin{equation}
\text{RMSE} = \sqrt{\frac{1}{n} \sum_{i=1}^{n} (y_i - \hat{y}_i)^2},
\end{equation}
where $y_i$ is the actual value, $\hat{y}_i$ is the predicted value, and $n$ is the number of data points.

\begin{figure} 
\includegraphics{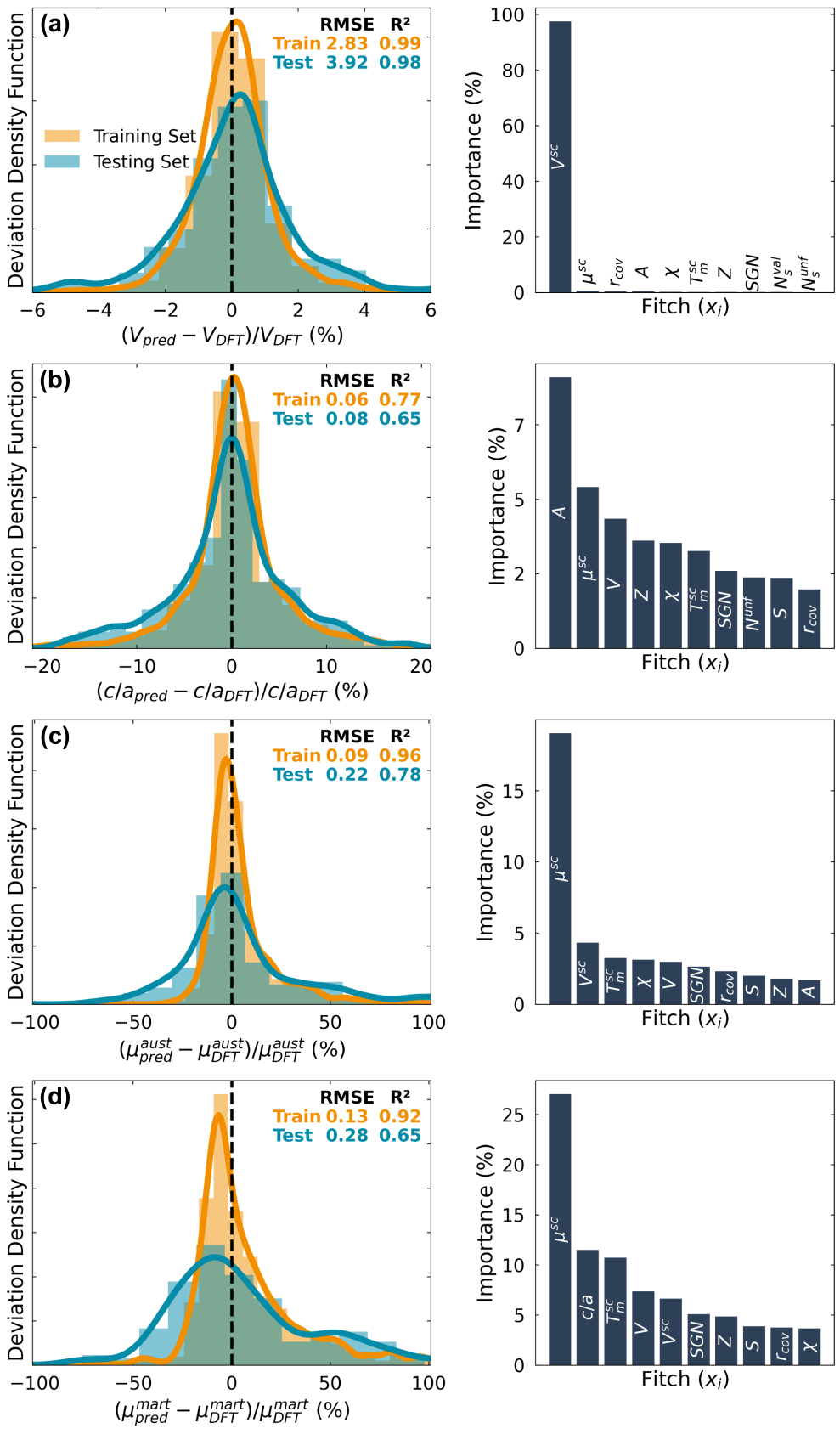} 
\caption{Distribution density of the regression prediction results with the $R^2$ and $RMSE$ metrics and feature importance for the training and test sets for (a) cell volume, (b) tetragonal ratio, and (c, d) magnetic moment of the austenitic and martensitic phases of all-$d$-metal metal Heusler alloys.} \label{deviation_density_functions} 
\end{figure}

A multi-objective genetic algorithm~\cite{murata1995moga} was employed to optimize the composition of all-$d$-metal Heusler alloys. The process began with the generation of multiple subpopulations, each comprising compositions situated at randomly selected points. The suitability of a composition was determined by the absolute difference in magnetizations predicted by the regression model, with a larger difference indicating greater suitability. Tournament selection with a fixed tournament size was used as the selection operator. To generate new subpopulations from the selected compositions, a crossover operator was employed to compute the arithmetic mean concentration of parental compositions. The mutation operator introduced random perturbations to the alloy compositions, ensuring that the overall composition was preserved, i.e., the sum of the element fractions equaled one. This perturbation involved transferring up to 5\% of one element's concentration to another element's concentration. The mutation and crossover rates were adjusted based on the diversity of the subpopulations, measured as the average pairwise distance between compositions in barycentric coordinates. At higher diversity, the rates were decreased to promote stability, while at lower diversity, the rates were increased to enhance the search.

\begin{figure} 
\includegraphics[width=0.47\textwidth]{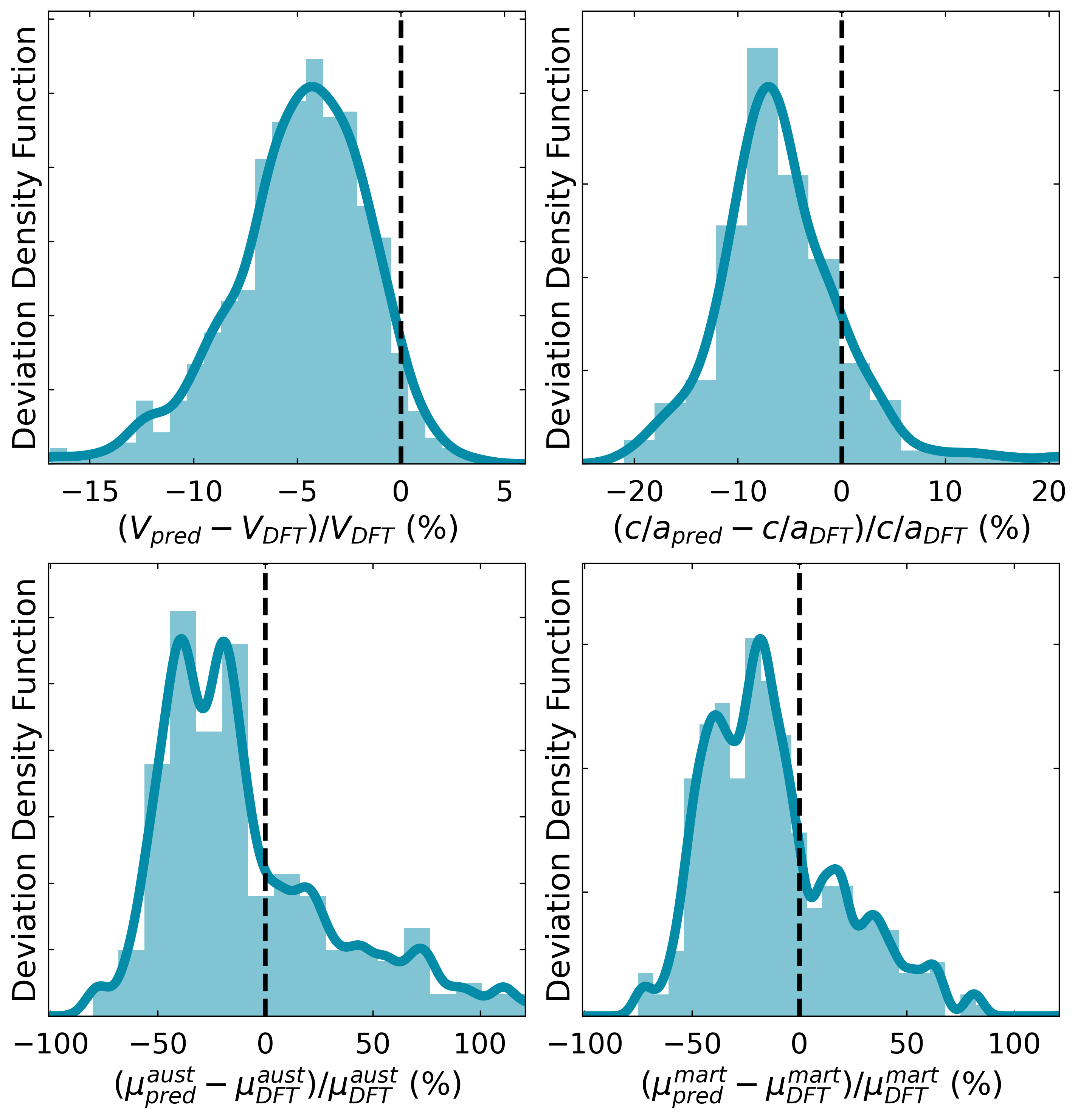} \caption{Density distribution of regression prediction error results relative to DFT calculations from Ref.~\cite{marathe2023exploration}.} 
\label{error_distributions}
\end{figure}

To validate the effectiveness of the regression model and the genetic algorithm, DFT was used to ensure that the computed minima accurately represent the physical properties of the system under investigation. This approach was applied specifically to the Ni-Co-Mn-Ti system. Initially, a series of compositions with an increment of 12.5\% was generated using 16 atomic supercells. Both austenitic and martensitic phases, including their direct and inverse structures, were constructed. Preliminary lattice parameters, including the initial volume and the c/a ratio, were derived from the regression model and further optimized through DFT calculations.

\begin{figure*}
\includegraphics[width=0.80\textwidth]{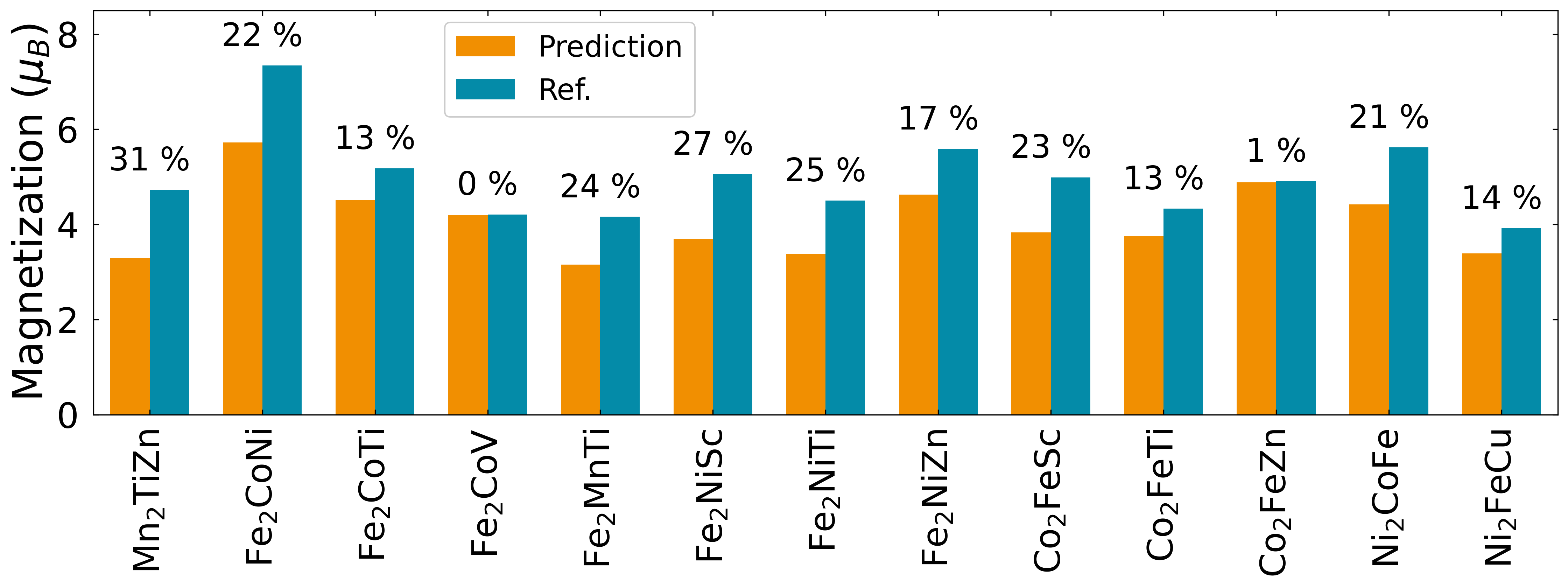} \caption{Magnetization for several all-$d$-metal Heusler alloys in the ground state. Reference values (Ref.) correspond to the values from paper~\cite{marathe2023exploration}. The numbers above the bars indicate the deviations between the magnetization obtained from DFT and the regression model.} 
\end{figure*}

\begin{figure}[b!] 
\includegraphics[width=0.47\textwidth]{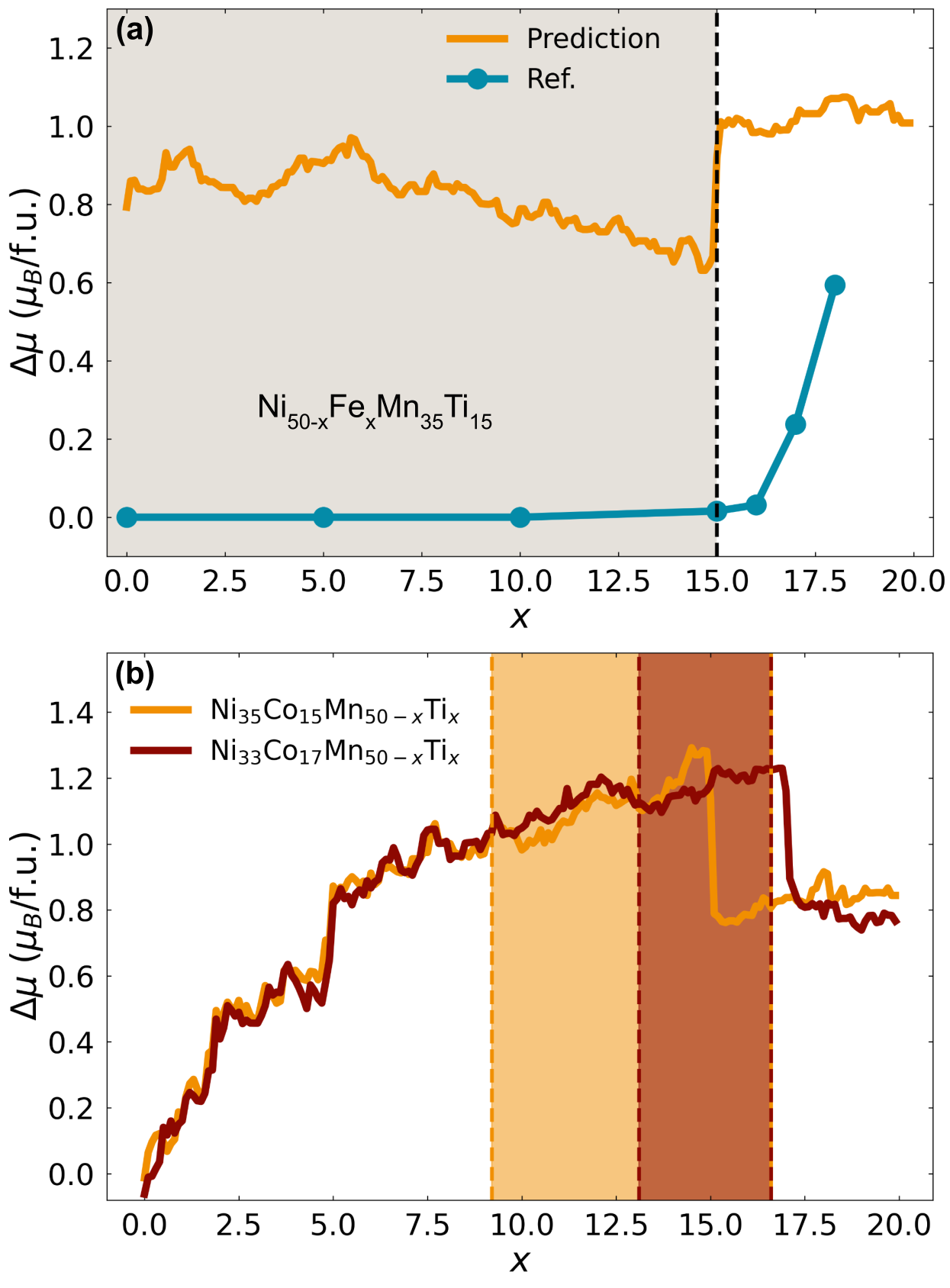} \caption{The difference in magnetization between the austenitic and martensitic phases $\Delta \mu$ for (a) Ni$_{50-x}$Fe$_x$Mn$_{35}$Ti$_15$ and (b) Ni$_{35}$Co$_{15}$Mn$_{50-x}$Ti$_x$ and Ni$_{33}$Co$_{17}$Mn$_{50-x}$Ti$_{x}$, depending on the concentration of the additional element. The experimental values are taken from the works~\cite{zeng2019electronic, beckmann2023dissipation}. Color areas indicate areas where the martensitic transition is possible.} 
\label{martensite_transition}
\end{figure}

To account for various magnetic orderings, a selection of both ferromagnetic (FM) and antiferromagnetic (AFM) structures was made. The AFM configurations were generated based on the system's symmetry using enumlib~\cite{morgan2017generating}, focusing on those with the highest symmetry, as these are often more energetically favorable. The selection process was thus limited to the five most symmetric AFM structures to reduce computational complexity. A comparison of the minima identified through the genetic algorithm and regression model with those obtained from DFT calculations was conducted to evaluate the reliability of the computational methods used in predicting the most promising configurations of the Ni-Co-Mn-Ti system.

\section{Results and discussion}

\subsection{Evaluation of the regression models}

Fig.~\ref{deviation_density_functions} illustrates the distribution density of the regression prediction results, alongside the $R^2$ and RMSE metrics for both the training and test sets, focusing on cell volume, tetragonality parameter, and total magnetic moment for the austenitic and martensitic phases. The model fully captures the cell volume, which represents an average value between the volumes of the elementary cells of its constituent elements, aligning well with the results of paper~\cite{li2023theoretical}. The dominant influence on this prediction is the average volume, with unaccounted percentages likely due to structural disorder.

\begin{figure*} 
\includegraphics[width=0.99\textwidth]{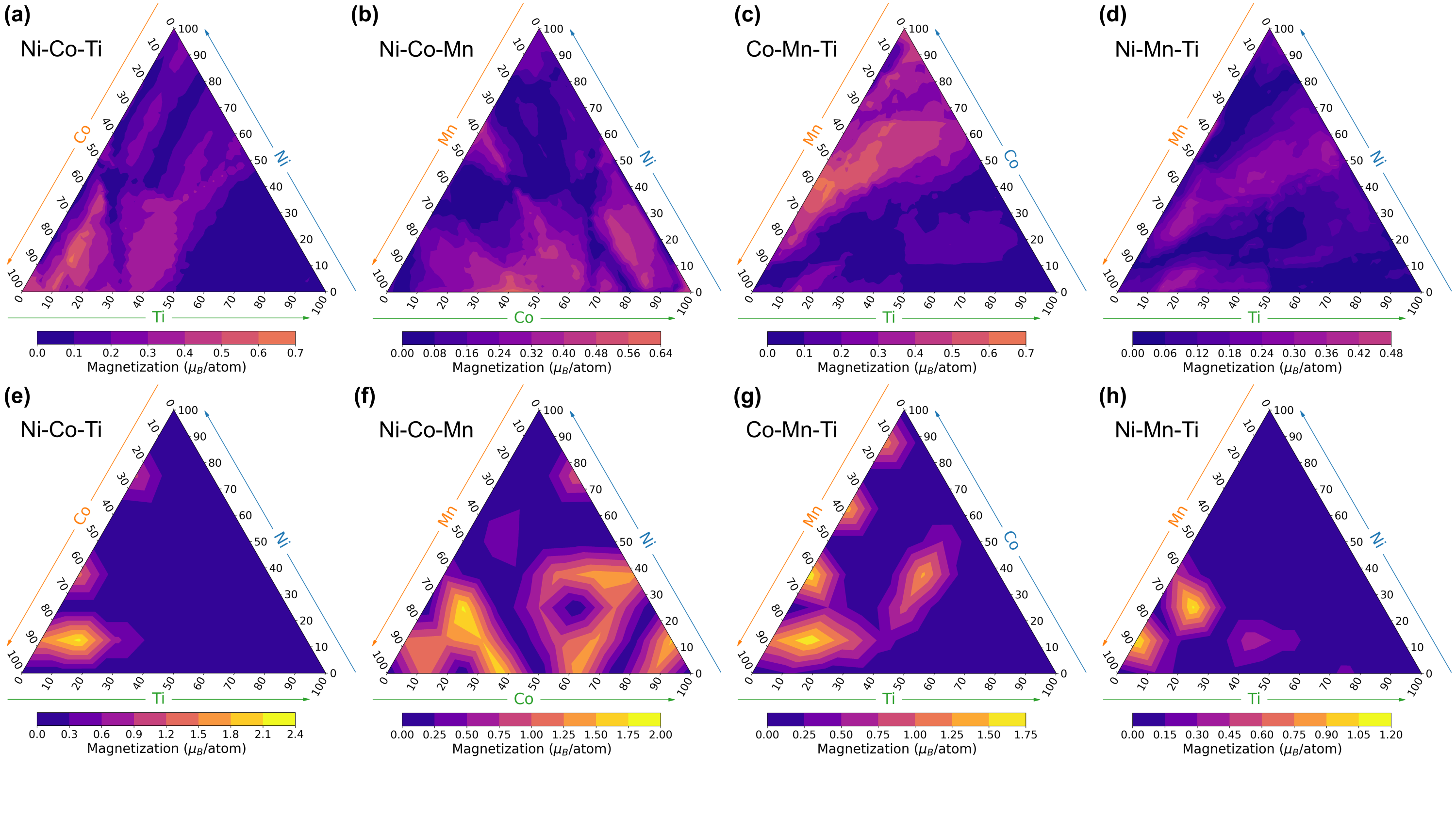} 
\caption{Ternary diagrams of the magnetization difference between austenite and martensite $\Delta \mu$ predicted by the regression model (upper panel) and calculated by DFT (lower panel) for (a, e) Ni-Co-Ti, (b, f) Ni-Co-Mn, (c, g) Co-Mn-Ti, (d, h) Ni-Mn-Ti.} 
\label{ternary_comparison}
\end{figure*}

\begin{figure*} 
\includegraphics[width=0.99\textwidth]{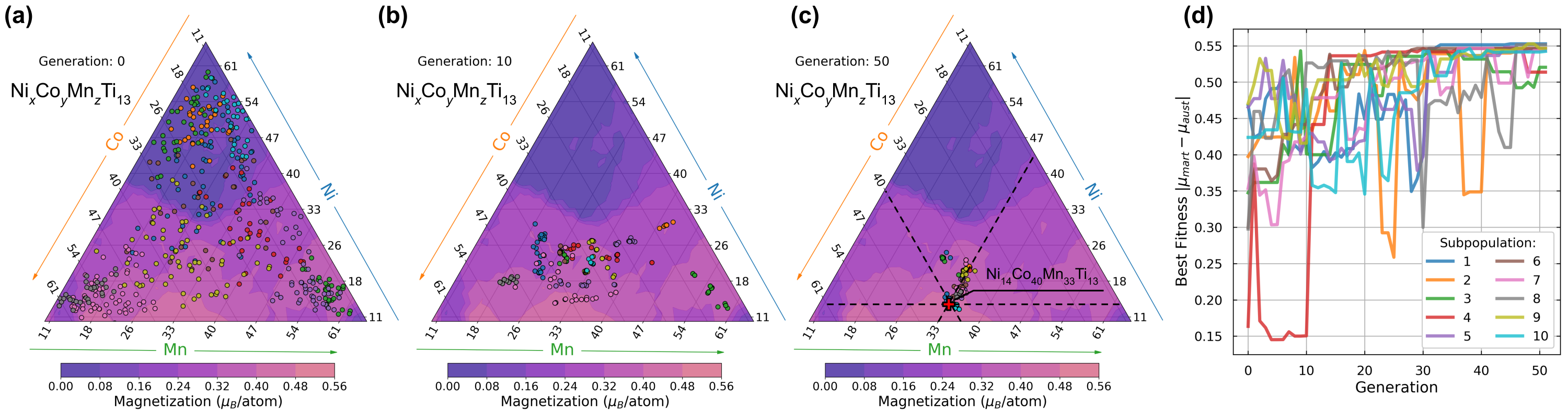} 
\caption{The results of the genetic optimization algorithm are presented in the form of ternary diagrams (a-c) showing the compositions of Ni$_x$Co$_y$Mn$_z$Ti$_{13}$ alloys, which were optimized to maximize the difference in magnetization between the austenitic and martensitic phases. Panel (d) shows the changes in the fitness function as a function of the number of steps for each subpopulation.} 
\label{ternary_genetic}
\end{figure*}

This approach is particularly effective for predicting volume, which mainly depends on the volume of its constituent components and can be approximated linearly. In all-$d$-metal Heusler alloys, the bonds are almost entirely metallic, and their volumes generally follow Vegard's law. The observed deviations from this law and the scattering of data points can be attributed to contributions from the magnetic subsystem, the presence of small covalent bonds, and other factors.
The tetragonality parameter is primarily influenced by the atomic mass of the alloy ($\approx$ 8\%), the magnetic moment of the constituent elements in their ground states ($\approx$ 5\%), and the cell volume ($\approx$ 4\%). The absolute error in predicting the tetragonality parameter is about 0.07, allowing for reliable predictions of the geometry of the martensitic phase in all-metal Heuslers, particularly those with a large $c/a$ ratio. It should be noted that this approach differs from the behavior observed in classical Heusler alloys, where the number of valence $d$ electrons is the primary determinant of tetragonal ratio~\cite{faleev2017heusler}. In the case of all-$d$-metal Heusler alloys, however, this contribution is significantly reduced, accounting for only $\approx$ 2\% of the overall effect.
For predicting magnetic moments in the case of austenite, the primary influence ($\approx$ 20\%) is the magnetic moment of the constituent elements in their ground state. The contribution from the following fitches is substantially less. The martensitic structure exhibits a similar trend, with the tetragonal ratio also having a significant impact ($\approx$ 12\%). The $R^2$ and RMSE values for magnetic moments show more pronounced differences between the test and training samples, indicating an increased risk of overfitting. Furthermore, the distribution profile becomes asymmetric, with a bias toward higher model estimates relative to DFT. These factors are likely due to the predominant presence of FM phases in the training set, whereas other magnetic states, such as AFM, are known to exist experimentally~\cite{guan2021first}.

To gain a more comprehensive understanding, it is also necessary to validate the model against data available in the literature. This approach helps identify and account for potential sources of deviation, thereby improving prediction accuracy. Initially, the prediction results obtained using the regression model were compared with the DFT calculation results from a previous study~\cite{fortunato2024high}. The deviation distribution based on this work approximately corresponds to the distribution shown in Fig.~\ref{deviation_density_functions}. However, there is a shift in the mean expectation to the left, likely due to the presence of 4$d$ and 5$d$ Heusler alloys in the validation dataset that were absent in the model's training dataset. This shift underscores the need to incorporate new data types into the model to enhance prediction accuracy. Nevertheless, the agreement between the model and the DFT results remains within an acceptable range, confirming its overall reliability and applicability.

The magnetic moments of the ground state for several all-$d$-metal Heusler alloys were predicted. Fig.~\ref{error_distributions} displays the predicted magnetic moments in comparison with those calculated in paper~\cite{marathe2023exploration}. The values obtained from the regression model align with the trends calculated using DFT. The maximum deviation in magnetic moment is 33\%, while the average deviation from the DFT values is 15\%, which is consistent with the results shown in Fig.~\ref{deviation_density_functions}. The largest deviation is observed for Mn$_2$TiZn. This discrepancy is likely due to the complex magnetic structure of this all-$d$-metal alloy, which includes FM, AFM, and non-magnetic configurations depending on the concentration of the constituent elements. These configurations are not sufficiently represented in the training dataset.

An indirect indicator of the magnetocaloric effect magnitude can be the difference in magnetic moments between the austenitic and martensitic phases, denoted as $\Delta \mu$. Fig.~\ref{martensite_transition} shows $\Delta \mu$ for the Ni-Mn-Ti system with Fe and Co substitutions. According to experimental data~\cite{zeng2019electronic}, no martensitic transition occurs up to a Fe concentration of 15\% in Ni$_{50-x}$Fe$_x$Mn$_{35}$Ti$_{15}$. Beyond this concentration, martensitic transformation becomes possible, and the magnetization difference between the austenitic and martensitic phases increases with higher Fe concentrations. The regression model results presented in Fig.~\ref{martensite_transition}(a) qualitatively replicate these findings, indicating a pronounced increase in magnetization at a Fe concentration of 15\%. However, the model currently cannot predict the complete absence of the martensitic phase at Fe concentrations up to 15\%. The results shown in Fig.~\ref{martensite_transition}(b) demonstrate similar trends for the Ni$_{35}$Co$_{15}$Mn$_{50-x}$Ti$_x$ and Ni$_{33}$Co$_{17}$Mn$_{50-x}$Ti$_x$ Heusler alloy series. The other authors experimentally and theoretically predicted that the martensitic transformation would cease at both low and high Ti concentrations~\cite{beckmann2023dissipation}, marked by colored areas at the boundaries of the martensitic transition.

The regression model can quantitatively predict the cessation of the transition at high concentrations, which is indicated by a sharp decline in $\Delta \mu$. At low Ti concentrations, a reduction in the magnetization difference is also observed, although the concentrations deviate from those observed experimentally. The absence of a martensitic transition is predicted for the Ni-Co-Mn system.

\subsection{Screening of the Ni-Co-Mn-Ti Alloy Family Using a Genetic Algorithm}

It is crucial to verify whether the maxima of magnetization differences correspond to those calculated by DFT. The Fig.~\ref{ternary_comparison} shows cross-sections of the Ni-Co-Mn-Ti diagram obtained using the regression model (a-d) and DFT calculations (e-h). In the case of the Ni-Co-Ti system, there is excellent agreement between the model results and the DFT data. For the Ni-Co-Mn and Co-Mn-Ti systems, good agreement is also observed, although the model does not identify all maxima. The most significant discrepancies are evident in the Ni-Mn-Ti system, where the model exhibits substantial deviations and predicts non-existent maxima. These errors are particularly pronounced in regions with high Mn concentrations and stabilization of the AFM state. This suggests that the current model may struggle to accurately predict magneto-structural states at high Mn concentrations and in scenarios dominated by complex magnetic interactions.

Optimization was performed for the Ni-Mn-Co-Ti system using this genetic algorithm with regression model. The results are shown on Fig.~\ref{ternary_genetic}. Initially, all compositions were distributed randomly, with each population centered around its focus. At the fifth step, no individuals remained in the zero magnetization region and the algorithm successfully finds local minima. The populations then migrate to the global maximum and successfully locate the region with the largest magnetization difference. Thus, the model is capable of identifying both local and global minima. The magnetization difference at each generation is shown in Fig.~\ref{ternary_genetic}(d). It can be seen that the curve oscillates with an upward trend. Oscillations correspond to the mutation moments to explore new regions. The best composition found in the Ni-Mn-Co-Ti class, according to the optimization results with the regression model, is Ni$_{15}$Co$_{39}$Mn$_{33}$Ti$_{13}$ with magnetization difference between austenitic and martensitic phases $\approx$ 2.24 $\mu_B$/f.u. This result aligns with physical intuition and findings from other studies~\cite{samanta2022large, zhang2022second}, since replacing weakly magnetic Ni and Ti with Co and Mn leads to increased magnetization.

\section{Conclusions}

In this study, a regression model based on the Random Forest algorithm was developed to predict the structural and magnetic properties of all-$d$-metal Heusler alloys. The model exhibits high accuracy in predicting structural characteristics such as cell volume and tetragonality parameter, while demonstrating moderate accuracy in predicting total magnetic moments of austenitic and martensitic phases.
The limited accuracy in predicting magnetic moments can be attributed to the inherent difficulty in determining AFM ground states. The model currently struggles to accurately capture the complex magnetic interactions that occur in systems with significant AFM contributions. Therefore, the development of a more effective and computationally affordable algorithm to accurately identify the ground magnetic state remains a critical challenge for improving the predictive capabilities of the model.

Despite these limitations, the model qualitatively predicts the presence or absence of martensitic transitions and the associated differences in magnetization between phases with a reasonable degree of accuracy. This capability is particularly important for applications where the martensitic transition plays a crucial role in determining material properties, such as in magnetocaloric materials.

Furthermore, a genetic optimization algorithm was employed to identify alloy compositions in family Ni-Co-Mn-Ti that exhibit the greatest differences in magnetization during the martensitic transition. These predicted compositions could be highly advantageous for magnetocaloric applications, where large magnetization changes are necessary. The results show a correlation with results from previous studies, underscoring the potential of this combined approach for discovering new materials with enhanced magnetocaloric properties.

\begin{acknowledgments}
The research was supported by the RSF - Russian Science Foundation project No. 24-12-20016. 
\end{acknowledgments}

\section*{Data Availability Statement}
The data that support the findings of this study, including a dataset of all-$d$-metal Heusler compounds, trained regression models, code for training and testing these models, and code for the optimization model, are openly available in the GitHub repository at the following link: \href{https://github.com/Danil-phy-cmp-120/all_d_optimization}{https://github.com/Danil-phy-cmp-120/all\_d\_optimization}.

\nocite{*}
\bibliography{bibliography}

\end{document}